\documentclass[aps,pra,reprint,superscriptaddress,nofootinbib,floatfix]{revtex4-2}

\usepackage{amsmath,amssymb,bm}
\usepackage{booktabs}
\usepackage{microtype}
\usepackage{xcolor}
\usepackage[colorlinks=true,linkcolor=blue,citecolor=blue,urlcolor=blue]{hyperref}
\hypersetup{
  pdftitle={DA-CASE: reusable measurements for adaptive quantum subspaces},
  pdfauthor={Ginanjar Utama and Hermawan Kresno Dipojono},
  pdfsubject={Single-reference adaptive subspace eigensolving with
    symmetry-aware selection and explicit measurement accounting}
}

\newcommand{\Tr}{\operatorname{Tr}}

\newcommand{\ket}[1]{\lvert #1\rangle}
\newcommand{\bra}[1]{\langle #1\rvert}

\newcommand{\supp}{\operatorname{supp}}
\newcommand{\rank}{\operatorname{rank}}
\newcommand{\mHa}{\mathrm{mHa}}

\begin{document}

\title{DA-CASE: reusable measurements for adaptive quantum subspaces}

\author{Ginanjar Utama}
\email{ginanjar.utama@gmail.com}
\affiliation{Department of Engineering Physics, Institut Teknologi Bandung,
Bandung, Indonesia}

\author{Hermawan Kresno Dipojono}
\email{dipojono@itb.ac.id}
\affiliation{Department of Engineering Physics, Institut Teknologi Bandung,
Bandung, Indonesia}

\date{August 9, 2026}

\begin{abstract}
Quantum subspace methods are often compared by basis dimension or energy
error, although their dominant experimental costs arise from different state
preparations, measurement settings, and shot allocations.  We present the
Dyadic Adaptive Clifford-Algebra Subspace Eigensolver (DA-CASE), whose basis
states are virtual directions \(A_i\ket{\psi}\) generated from one reference.
Overlap, Hamiltonian, and observable matrices are reconstructed from one
cached set of Pauli expectations on that reference.  The method therefore
trades multiple prepared basis states for a potentially wide measurement bank.
We make that trade explicit on a frozen eight-qubit H$_4$ Hamiltonian.  Two
generator resolutions reach the same nine-dimensional subspace and the same
energy to machine precision, while the retained bank changes from \(7371\) to
\(2240\) Pauli words.  A reference-conditioned symmetry test certifies the
narrower span without asserting that its individual Pauli words conserve the
sector as abstract operators.  Independently, a dyadic commuting hierarchy
reduces the determinant bank from \(913\) qubit-wise-commuting settings to
\(64\) fully commuting settings, while exposing the added logical-CX cost.
In a separate four-qubit finite-shot diagnostic, covariance-aware allocation
reduces the projected-matrix variance target by \(68.9\%\).  Mode-wise overlap
regularization removes the observed catastrophic energy estimates and lowers
RMSE, but doubles the median error relative to a fixed cutoff.  These are
small-instance exact and Monte Carlo results, not a hardware demonstration,
scaling result, or quantum advantage claim.  The contribution is a
single-reference measurement architecture and a resource ledger that keeps
contexts, settings, shots, circuit depth, and post-selection retries in their
proper units.
\end{abstract}

\maketitle

\section{Introduction}
\label{sec:introduction}

Quantum subspace diagonalization replaces a deep variational search by the
generalized eigenproblem
\begin{equation}
  H_{\mathrm{sub}}c = E S c ,
  \label{eq:gep}
\end{equation}
in a small, generally nonorthogonal basis.  Quantum subspace expansion,
nonorthogonal eigensolvers, and Krylov methods differ mainly in how that basis
is constructed and how its matrix elements are measured
\cite{mcclean2017qse,huggins2020nonorthogonal,stair2020krylov,bharti2021iqae}.
The conditioning of \(S\) is central because finite-shot perturbations are
amplified near its null space
\cite{epperly2022theory,lee2024sampling,zhang2024measurement}.  Recent
Davidson, partitioned, and overlap-adaptive variants make basis construction
more flexible, but do not remove this measurement problem
\cite{tkachenko2024davidson,oleary2025partitioned,feniou2023overlapadapt,
miura2026assqd}.

The experimental primitive is a prepared-state measurement shot.  A basis
with few vectors can still require many state contexts, settings, or
off-diagonal transition measurements; conversely, one setting can return
several compatible Pauli expectations.  Raw basis size, state count, Pauli
word count, and circuit executions are therefore not interchangeable.  This
distinction is especially important when comparing subspace methods with
VQE and ADAPT-VQE \cite{peruzzo2014vqe,grimsley2019adapt}, or with
generating-function approaches such as ADAPT-GCIM \cite{zheng2024adaptgcim}.

We study a deliberately specific architecture.  The Adaptive
Clifford-Algebra Subspace Eigensolver (A-CASE) represents every direction as an
operator \(A_i\) acting virtually on one reference.  The Dyadic Adaptive
Clifford-Algebra Subspace Eigensolver (DA-CASE) adds a hierarchy of
quantum-information Clifford measurement circuits.  The two uses of
Clifford are distinct: the former is an operator representation, while the
latter is stabilizer-circuit synthesis for commuting measurements.  This paper
asks one question:
\begin{quote}
What does a fixed-reference operator bank save, and what does it cost, when
all methods are expressed in measurement-compatible units?
\end{quote}

The answer has four parts.
\begin{enumerate}
 \item All projected matrices are linear reconstructions from one expectation
 bank on one reference.  This removes separately prepared basis states but can
 enlarge the Pauli universe.
 \item Generator resolution is a real resource variable.  On H$_4$, coarse
 determinant excitations and their Pauli-word resolution reach the same
 retained subspace with widths \(7371\) and \(2240\).
 \item A reference-aware sector test accepts state-safe Pauli directions that
 fail a stronger operator-global test, and certifies the entire retained span.
 \item Measurement grouping and overlap regularization expose separate
 depth--setting and bias--tail-risk trades; neither implies an end-to-end
 advantage.
\end{enumerate}

The intended input boundary is an effective many-body Hamiltonian, for example
one obtained after density-functional, Wannier, screening, or embedding steps
\cite{marzari2012wannier,pizzi2020wannier90,aryasetiawan2004crpa,
georges1996dmft}.  DA-CASE is not a DFT method and does not replace those
stages.  We use FCIDUMP as a strict interchange format
\cite{knowles1989fcidump}, with PySCF and OpenFermion providing independent
construction checks \cite{sun2018pyscf,mcclean2020openfermion}.

\section{Method}
\label{sec:method}

\subsection{One reference, one matrix-element bank}
\label{sec:bank}

Let \(H=\sum_w h_w P_w\) be an \(n\)-qubit Hamiltonian and
\(\rho=\ket{\psi}\bra{\psi}\) a normalized reference.  For an ordered
generator family \(\{A_i\}_{i=0}^{M-1}\), with \(A_0=I\), define
\begin{align}
 S_{ij} &= \Tr(\rho A_i^\dagger A_j), \label{eq:s}\\
 H_{ij} &= \Tr(\rho A_i^\dagger H A_j), \label{eq:h}\\
 Q_{ij} &= \Tr(\rho A_i^\dagger Q A_j). \label{eq:q}
\end{align}
Every product is expanded in the Hermitian Pauli basis.  Complex expansion
coefficients are permitted, but every measured Pauli expectation is real.
Consequently, all entries are reconstructed from expectations on the same
\(\rho\); no Hadamard test or separately prepared \(A_j\ket{\psi}\) is used.

For a retained basis \(\mathcal B\), the required word universe is
\begin{align}
 \mathcal W(\mathcal B)
 &=
 \bigcup_{i,j\in\mathcal B}\supp(A_i^\dagger A_j)
 \nonumber\\
 &\quad\cup
 \bigcup_{i,j\in\mathcal B}\supp(A_i^\dagger H A_j),
 \qquad W=|\mathcal W|.
 \label{eq:word-universe}
\end{align}
Projected observables extend the same cache by
\(\supp(A_i^\dagger Q A_j)\).  If compatible words are partitioned into
\(G\) settings and setting \(g\) receives \(n_g\) shots, the physical
preparation count is \(\sum_g n_g\), not \(W\) or \(G\).  One shot supplies
the joint outcome needed for every compatible word in its setting.

We therefore report
\begin{equation}
 (M,\rank S,\kappa_S,W,G,C_{\mathrm{ctx}},D_{\mathrm{prep}}),
 \label{eq:ledger}
\end{equation}
where \(C_{\mathrm{ctx}}\) counts distinct state or parameter contexts and
\(D_{\mathrm{prep}}\) records state-preparation depth or rotors.  Logical
measurement-circuit gates, shot allocation, and post-selection retries are
separate fields.  This separation is the main accounting rule of the paper.

\subsection{Adaptive growth and generator resolution}
\label{sec:growth}

Given a normalized Ritz pair \((E,\ket{\Psi})\) and a candidate
\(\ket{\chi_a}=A_a\ket{\psi}\), A-CASE evaluates the exact generalized
\(2\times2\) pencil
\begin{equation}
 \begin{pmatrix}E&h_a\\h_a^*&h_{aa}\end{pmatrix}v
 =\lambda
 \begin{pmatrix}1&s_a\\s_a^*&s_{aa}\end{pmatrix}v .
 \label{eq:two-by-two}
\end{equation}
The score is the predicted lowering
\(\Delta_a=\max(0,E-\lambda_{\min})\), optionally penalized by the number of
new words introduced.  Candidates that annihilate the reference or have too
little component outside the retained span are rejected.  The selected
direction appends one cached row and column before the full pencil is solved.

A determinant excitation is a Pauli sum, not an indivisible object.  A-CASE
may therefore score the complete excitation or its component Pauli words.
The two resolutions need not select the same route, but if they reach the same
span their energies and observables coincide while their word universes may
differ.  This is tested directly rather than assumed.

\subsection{Reference-aware symmetry}
\label{sec:symmetry}

An operator-global leakage test first checks commutators with particle number
and \(S_z\).  It is sufficient but stronger than the condition needed by a
fixed reference: an individual Pauli word can fail globally while
\(A_a\ket{\psi}\) remains in the declared sector.  For a target projector
\(P_q\), \(q=(N,S_z)\), the reference-conditioned leakage is
\begin{equation}
 \ell_a^{\mathrm{ref}}
 =1-
 \frac{\Tr(\rho A_a^\dagger P_q A_a)}
      {\Tr(\rho A_a^\dagger A_a)} .
 \label{eq:reference-leakage}
\end{equation}
The spin-orbital ordering is an explicit input to \(S_z\), inference, and
projection.  This matters because a determinant is a sharp \(S_z\) eigenstate
under both interleaved and blocked conventions, but generally with different
eigenvalues.

Individual tests do not certify linear combinations.  The retained span is
therefore checked with
\begin{equation}
 L_{ij}=\Tr[\rho A_i^\dagger(1-P_q)A_j],
 \qquad
 \ell_{\mathcal B}=\lambda_{\max}(L,S).
 \label{eq:span-certificate}
\end{equation}
Then \(\ell_{\mathcal B}\le\tau\) bounds the sector leakage of every normalized
combination in the retained subspace.  The result is reference-conditioned;
it does not relabel an accepted Pauli word as a globally conserving operator.

A sector-mixed warm reference cannot be assigned a sharp sector by rounding an
expectation value.  It may instead be post-selected as
\begin{equation}
 \rho_q=\frac{P_q\rho P_q}{\Tr(P_q\rho)} .
 \label{eq:postselection}
\end{equation}
The acceptance probability \(p_q=\Tr(P_q\rho)\) gives an expected retry factor
\(1/p_q\) for subsequent A-CASE preparation executions.  It does not multiply
the number of distinct settings, and it does not apply retrospectively to the
ADAPT measurements used to construct the warm reference.  Realizing
Eq.~\eqref{eq:postselection} without resolving a determinant requires a
quantum non-demolition measurement of \(N\) and \(S_z\); its ancilla, gate,
and latency costs are not priced here.

\subsection{Dyadic commuting measurements}
\label{sec:dyadic}

Qubit-wise commuting (QWC) groups are shallow but can be numerous
\cite{verteletskyi2020grouping}.  For \(n=2^L\) qubits and block size
\(k=2^\ell\), partition the register into contiguous blocks \(B_{k,b}\).
Two Pauli words are \(k\)-compatible when
\begin{equation}
 [P|_{B_{k,b}},Q|_{B_{k,b}}]=0
 \quad\text{for every }B_{k,b}.
 \label{eq:block-commuting}
\end{equation}
The endpoints are QWC at \(k=1\) and full Pauli commutation at \(k=n\).
Each compatible group is diagonalized by block-local stabilizer Clifford
circuits \cite{crawford2021efficient,miller2024hardware}.  We verify that every
word, not only an independent stabilizer basis, maps to a \(Z\)-only word.
The coloring is constructive rather than optimal.  The hierarchy leaves the
subspace and \(W\) unchanged while trading fewer settings for more logical
CX gates and depth.

\subsection{Finite-shot assembly and regularization}
\label{sec:finite-shot-method}

For every QWC group, the implementation stores its joint shot histogram.
Means, covariances, projected matrices, and bootstrap replicas are derived
from that shared object.  Grouped resampling follows the nonparametric
bootstrap principle \cite{efron1979bootstrap} and preserves within-setting
covariance.

The measured overlap \(\hat S\) is normalized on live generators:
\(\bar S=D^{-1}\hat S D^{-1}\), with
\(D_{ii}=\sqrt{\hat S_{ii}}\) and rows below a declared norm floor removed.
Let \((\lambda_k,u_k)\) be the canonically ordered eigenpairs of \(\bar S\).
A covariance-propagated radius \(r_k\) is computed for
\(u_k^\dagger\bar S u_k\), and mode \(k\) is retained when
\begin{equation}
 \lambda_k >
 \max\!\left(\tau_S,\frac{\lambda_{\max}}{\kappa_{\max}},r_k\right).
 \label{eq:mode-cutoff}
\end{equation}
Using the same maximum radius for every mode is retained as a uniform
comparator.  Because \(D\), \(u_k\), and the rank are selected from the same
data used in the solve, this rule is a heuristic regularizer, not a confidence
or variational certificate.

For shot allocation, a pilot cache supplies group variances for a family of
matrix-element functionals and one first-order Ritz functional.  If the
linearized variance is \(\sum_g V_g/n_g\), the Neyman allocation is
\begin{equation}
 n_g^* \propto \sqrt{V_g}.
 \label{eq:neyman}
\end{equation}
This optimizes the declared local variance target; it does not by itself
optimize the nonlinear energy estimator.

\section{Evaluation}
\label{sec:evaluation}

\subsection{Primary \texorpdfstring{H$_4$}{H4} contract}
\label{sec:h4-contract}

The main comparison uses a frozen eight-qubit H$_4$ FCIDUMP at \(0.9\) \AA,
four electrons, and \(S_z=0\).  Its file digest, geometry, basis, RHF energy,
and determinant-FCI reference are recorded.  The mapped sector energy agrees
with the external reference to \(3.1\times10^{-15}\) Ha, and an independent
OpenFermion reconstruction agrees over \(185\) Pauli terms to
\(8.74\times10^{-16}\).

All adaptive subspace arms use a nine-direction target where applicable.
The matched table reports exact workflow contexts, selection work, retained
width, and QWC groups.  These are structural counts.  At uniform \(R\) shots
per setting, a setting count becomes \(R\) preparation executions; no uniform
\(R\) is assumed to deliver equal precision across methods.

\begin{table*}[t]
\begingroup
\setlength{\tabcolsep}{2pt}
\caption{Matched H$_4$ contract.  Ctx.\ counts state or parameter contexts,
not physical preparations.  Sel.\ \(W/G_\Sigma\) reports distinct selection
words and QWC setting evaluations summed over changing-state steps.  Retained
\(W/G\) reports the final bank.  \(H/S\) lists transition-matrix pairs for
ADAPT-GCIM, whose measurement primitive is not a single-reference word bank.
The two warm-start rows have the same setting counts because post-selection
changes the prepared state and retry rate, not the measured settings.}
\label{tab:matched}
\scriptsize
\begin{ruledtabular}
\begin{tabular}{lccccccccc}
Arm & \(M\) & Error (\(\mHa\)) & \(\kappa_S\) & Ctx. & Rot. & Sel. & Sel.\ \(W/G_\Sigma\) & \(W/G\) & \(H/S\) pairs \\
\colrule
QSE & 9 & 41.653 & 1 & 1 & 0 & --- & --- & 4,226/921 & --- \\
Krylov & 9 & $1.05\times10^{-5}$ & $6.6\times10^{10}$ & 1 & 0 & --- & --- & 4,224/---\footnotemark[1] & --- \\
generator coordinate & 9 & 49.576 & 1 & 1 & 0 & --- & --- & 13,646/1,705 & --- \\
ADAPT-GCIM (4 iter., M=8) & 8 & 13.364 & $3.3\times10^{1}$ & 8 & 4 & 98 & 2,424/1,812 & n/t & 36/28 \\
ADAPT-GCIM (8 iter., M=16) & 16 & 10.674 & $1.1\times10^{3}$ & 16 & 8 & 180 & 2,424/3,572 & n/t & 136/120 \\
ADAPT-VQE & --- & 2.375 & --- & 90 & 8 & 1,252 & 2,424/3,616 & 185/68 & --- \\
DA-CASE (determinant) & 9 & 3.019 & 1 & 1 & 0 & 180 & 15,783/1,681 & 7,371/913 & --- \\
DA-CASE (word, leakage rejected) & 1 & 56.057 & 1 & 1 & 0 & --- & 185/68 & 185/68 & --- \\
DA-CASE (word) & 9 & 3.019 & 1 & 1 & 0 & 1,252 & 14,401/1,672 & 2,240/465 & --- \\
DA-CASE (det., ADAPT warm start) & 9 & 0.342 & 1.02 & 18 & 2 & 499 & 15,803/2,589 & 7,510/913 & --- \\
DA-CASE (det., sector-projected ADAPT warm start) & 9 & 0.199 & 1.02 & 18 & 2 & 499 & 15,803/2,589 & 7,510/913 & --- \\

\end{tabular}
\footnotetext[1]{The Krylov word count comes from a separately certified
support calculation; QWC grouping was not computed.}
\end{ruledtabular}
\endgroup
\end{table*}

\subsection{Auxiliary tests and evidence labels}
\label{sec:auxiliary}

The broader suite includes H$_2$, LiH, equilibrium and stretched H$_4$,
two H$_2$O active spaces, \(2\times2\) and \(2\times3\) Hubbard clusters,
a \(2\times2\) Kitaev cluster, and a two-site Hubbard dimer.  These systems
check matrix assembly, generator families, response reconstruction, and
conditioning; they do not establish scaling.

Evidence labels are carried in the records.  Noiseless projected solves are
\emph{exact} for their declared subspaces.  Exact-ground-state sampling used
to test QSCI-style configuration selection is labelled
\emph{oracle sampled}, following the distinction emphasized in recent QSCI
work \cite{nakagawa2024adaptqsci,kanno2026qsci,gaberle2026critical}.  Bootstrap
and overlap-calibration studies are labelled \emph{heuristic}.  Reported
interval summaries are conditional on the surviving replicas.  Response
reconstruction follows projected spectral methods
\cite{colless2018spectra,umeano2025response,patel2026qsense}; it is a secondary
reuse test rather than the paper's main claim.

\section{Results}
\label{sec:results}

\subsection{The bank changes the resource profile, not the variational target}
\label{sec:results-profile}

The matched H$_4$ record illustrates why state contexts and preparations must
be separated.  Cold A-CASE uses one reference context at \(M=9\).  ADAPT-GCIM
uses eight or sixteen generating-function states, and ADAPT-VQE visits ninety
selection or optimizer contexts.  None of these counts is a physical
preparation count: every shot at every setting requires a fresh preparation.

At the setting level, word-resolution A-CASE evaluates a \(14401\)-word
selection cache in \(1672\) QWC settings on one reference.  ADAPT-VQE's
\(2424\)-word union uses \(452\) settings at each of eight changing selection
states, giving \(3616\) setting evaluations before optimizer measurements.
The latter also has a lower bound of \(5576\) Hamiltonian setting evaluations
for its exact optimizer energy calls, while a physical gradient protocol
remains unspecified.  These numbers describe schedules, not equal-precision
performance.

Energy gives the complementary view.  Krylov is essentially exact on this
small instance but has \(\kappa_S=6.6\times10^{10}\).  ADAPT-VQE is more
accurate than the cold A-CASE arms.  The fixed-reference architecture is
therefore not justified by matched-\(M\) energy; its value is reuse and an
explicit measurement boundary.

\subsection{Generator resolution reduces width without changing the span}
\label{sec:results-resolution}

Determinant-resolution and word-resolution A-CASE both retain \(M=9\),
\(\kappa_S=1\), and an error of \(3.019\) mHa.  Their nine principal angles
are zero to numerical precision and their energies differ by
\(4\times10^{-16}\) Ha.  Nevertheless, the retained word universe falls from
\(7371\) to \(2240\), and the QWC group count from \(913\) to \(465\).
The same effect appears on the \(2\times2\) Hubbard plaquette, where the
retained width falls from \(2519\) to \(413\).

The saving is not free.  H$_4$ selection grows from \(180\) candidate
scorings at determinant resolution to \(1252\) at word resolution, and the
selection cache remains much wider than the retained bank.  Resolution moves
work from final measurement into selection.

Every odd-\(Y\) word in this pool fails the operator-global symmetry filter.
The reference-aware cascade accepts the directions that keep the declared H$_4$
reference in sector, and Eq.~\eqref{eq:span-certificate} returns zero leakage
for the full retained span.  The negative-control arm using only the global
filter remains at the identity and has \(56.057\) mHa error.  Thus the
reference-aware test enables the narrower representation without weakening
the meaning of the certificate.

The warm-start comparison is intentionally separate.  The unprojected ADAPT
reference has sector weight \(0.999882\) and cannot receive a sharp-sector
certificate.  Algebraic post-selection gives unit sector weight and changes
the A-CASE error from \(0.342\) to \(0.199\) mHa.  Its acceptance probability
is \(0.999880\), corresponding to a \(1.00012\) expected retry factor on
subsequent A-CASE preparation executions.  The ADAPT prelude and all setting
counts are unchanged, and the non-demolition sector-measurement circuit is
unpriced.  This is a proof-of-principle symmetry operation, not a hardware
cost result.

\subsection{Dyadic grouping trades settings for entangling depth}
\label{sec:results-dyadic}

The dyadic hierarchy is applied to the retained determinant bank while holding
the subspace and \(W=7371\) fixed.  Table~\ref{tab:dyadic} shows a monotone
reduction from \(913\) QWC settings to \(64\) fully commuting settings.
At \(8000\) uniform shots per setting, preparation executions fall from
\(7.304\) million to \(0.512\) million.  The fully commuting endpoint requires
\(2006\) logical CX gates over one complete setting sweep, with mean and
maximum two-qubit depths \(25.72\) and \(42\).  Connectivity, routing, device
noise, and error mitigation are excluded.

\begin{table*}[t]
\caption{Constructive dyadic block-commuting hierarchy.  \(G\) is the number
of settings, \(N_{\rm CX}\) the logical CX count over one sweep, and Preps.\
assumes \(8000\) uniform shots per setting.  The last column is the logical
break-even ratio \(c_{\rm CX}/c_{\rm prep}\) relative to QWC; it is not a
device-runtime prediction.}
\label{tab:dyadic}
\scriptsize
\begin{ruledtabular}
\begin{tabular}{llrrrrrrr}
System & \(k\) & \(G\) & \(W/G\) & \(N_{\rm CX}\) & mean \(D_{\rm CX}\) & max \(D_{\rm CX}\) & Preps.\ (\(10^6\)) & \(c_{\rm CX}/c_{\rm prep}\) \\
\colrule
H$_4$ ($M=9$, $W=7,371$) & $1$ (QWC) & 913 & 8.07 & 0 & 0.00 & 0 & 7.304 & --- \\
 & $2$ & 647 & 11.39 & 3185 & 2.55 & 4 & 5.176 & 0.084 \\
 & $4$ & 238 & 30.97 & 3688 & 8.90 & 17 & 1.904 & 0.183 \\
 & $8$ (full) & 64 & 115.17 & 2006 & 25.72 & 42 & 0.512 & 0.423 \\
\colrule
BeH$_2$ ($M=5$, $W=1,815$) & $1$ (QWC) & 353 & 5.14 & 0 & 0.00 & 0 & 2.824 & --- \\
 & $2$ & 41 & 44.27 & 168 & 1.78 & 2 & 0.328 & 1.857 \\
 & $4$ & 26 & 69.81 & 234 & 5.12 & 10 & 0.208 & 1.397 \\
 & $8$ (full) & 14 & 129.64 & 332 & 19.00 & 27 & 0.112 & 1.021 \\

\end{tabular}
\end{ruledtabular}
\end{table*}

The hierarchy demonstrates a tunable logical trade, not a preferred block
size.  A hardware choice requires the routed circuit, gate errors, readout,
reset, and workload-specific precision.  Counting fewer settings while
ignoring their deeper measurement circuits would simply move an omitted cost.

\subsection{Finite-shot stabilization trades median accuracy for tail control}
\label{sec:results-finite-shot}

The finite-shot study fixes one four-qubit TFIM bank with \(M=9\), exact rank
eight, \(\kappa_S=194.94\), \(189\) words, and \(52\) QWC groups.  Each of
\(200\) replicas receives \(104000\) physical setting-shots.  Uniform
allocation uses \(2000\) shots per group.  Covariance-aware allocation uses a
\(200\)-shot pilot per group and distributes the remainder according to
Eq.~\eqref{eq:neyman}.

The adaptive allocation reduces the median summed projected-matrix variance
from \(272.46\) to \(84.75\), or \(68.9\%\), and the exact-Ritz first-order
variance by \(20.0\%\).  Under the fixed overlap cutoff, the median absolute
error barely changes, from \(4.48\) to \(4.44\) mHa, but catastrophic
\(>0.1\) Ha estimates fall from four to one.  Rare near-null modes dominate
RMSE.

Uniform calibration applies the largest radius to every overlap mode and
over-truncates this bank.  Mode-wise calibration recovers rank eight in
\(71/200\) uniform and \(67/200\) covariance-aware replicas.  With
covariance-aware allocation it gives \(8.73\) mHa median error, \(12.85\) mHa
RMSE, and no catastrophic replica, compared with \(4.44\) mHa median error,
\(285.41\) mHa RMSE, and one catastrophe for the fixed cutoff.

\begin{table*}[t]
\caption{Allocation and overlap regularization on the fixed TFIM bank.
Errors are relative to the exact projected energy.  The data-derived
mode-wise rule is a heuristic; \(>0.1\) Ha counts catastrophic replicas and
rank \(8\) counts recovery of the exact effective rank.}
\label{tab:finite-shot}
\scriptsize
\begin{ruledtabular}
\begin{tabular}{llccccccc}
Allocation & Overlap rule & median & p95 & max & RMSE & bias & \(>0.1\) Ha & rank \(8\) \\
\colrule
uniform & fixed & 4.48 & 37.01 & 22760.2 & 1786.21 & -178.08 & 4/200 & 200/200 \\
 & calibrated (uniform) & 21.40 & 90.17 & 97.8 & 50.14 & +38.15 & 0/200 & 0/200 \\
 & calibrated (per mode) & 10.64 & 23.96 & 34.8 & 13.31 & +10.33 & 0/200 & 71/200 \\
\colrule
covariance-aware & fixed & 4.44 & 30.79 & 4032.6 & 285.41 & -24.94 & 1/200 & 200/200 \\
 & calibrated (uniform) & 81.42 & 91.42 & 93.4 & 64.65 & +53.46 & 0/200 & 0/200 \\
 & calibrated (per mode) & 8.73 & 23.67 & 39.4 & 12.85 & +8.33 & 0/200 & 67/200 \\

\end{tabular}
\end{ruledtabular}
\end{table*}

The lower RMSE is not a variational statement.  The calibrated arms show a
positive mean bias of \(8\)--\(10\) mHa, consistent with truncation, but
\(22/200\) and \(34/200\) mode-wise replicas still lie below the exact energy.
Both the matrices and the retained rank are selected from noisy data.
Separating estimator scatter from a truncation floor would require an
exact-matrix truncated-rank control.  The present result supports a
bias--tail-risk choice on this bank only.

\section{Scope and limitations}
\label{sec:limits}

The results support a resource architecture, not a universal accuracy claim.
Four limits are decisive.

\begin{enumerate}
 \item \emph{Selection can dominate.}  The H$_4$ selection cache contains
 \(14401\)--\(15803\) words, much more than the retained bank.  Reuse across
 observables or repeated solves is needed to amortize that cost.
 \item \emph{Conditioning remains physical.}  Exact nested subspaces are
 variational, but noisy pencils are not.  In the auxiliary response test,
 increasing \(\kappa_S\) by four orders reduces successful bootstrap replicas
 from \(200/200\) to \(137/200\) and widens a susceptibility interval by about
 fifteen times.
 \item \emph{Sampling-fed selection is not an established gain.}  On the
 \(2\times3\) Hubbard test at \(M=7\), operator dressing is indistinguishable
 from a sample-independent selected-CI control.  Exact-ground-state samples
 are an oracle input, not an implementable preparation claim.
 \item \emph{No scaling result is shown.}  All systems admit exact classical
 validation.  Finite-shot adaptive growth at eight qubits remains costly and
 unreliable in the present implementation.
\end{enumerate}

These limitations prevent a total-resource comparison with ADAPT-VQE.  Its
physical gradient protocol and precision-matched group variances are not
specified, while DA-CASE's QND post-selection circuit is likewise unpriced.
The H$_4$ setting comparison is therefore a scheduling diagnostic and not a
total-resource advantage claim.  We make no quantum advantage claim.

\section{Discussion}
\label{sec:discussion}

The useful abstraction is not a scalar exchange rate between prepared states
and Pauli words.  For method \(X\), a device-level runtime begins with
\begin{equation}
 T_X=\sum_{s,g} n_{sg}
 \left[t_{\rm prep}^X(s)+t_{\rm basis}(g)+t_{\rm ro}+t_{\rm reset}\right],
 \label{eq:runtime}
\end{equation}
where \(s\) labels a state or parameter context and \(g\) a compatible
measurement setting.  A fixed-reference bank reduces the number of \(s\)
values; generator resolution changes the word and setting sets; dyadic
grouping changes \(t_{\rm basis}\); shot allocation changes \(n_{sg}\); and
post-selection multiplies only the affected preparation attempts.  The terms
belong in one execution model, but they must not be collapsed before their
hardware meanings are defined.

This view suggests three next steps.  First, freeze and reuse large selection
caches across related Hamiltonians or observables.  Second, test
reference-aware symmetry on sector-sharp multideterminant references that do
not require post-selection.  Third, route the dyadic diagonalizers onto a
declared topology and combine their depth with measured gate and readout noise.
Only then can settings, shots, and circuit time support a precision-matched
hardware comparison.

\section{Reproducibility}
\label{sec:reproducibility}

The implementation, frozen Hamiltonians, benchmark records, generators, tests,
and manuscript are maintained together in the private development repository
\texttt{clifford\_qc}.  Public access to the moving branch is not claimed.
Editors and referees can receive an access-controlled frozen snapshot with the
source revision, data, environment, and artifact generators.  A tagged
archival release with a persistent identifier is intended after the interfaces
and evidence contracts stabilize.

The H$_4$ interchange file is bound to its record by SHA-256.  Every table
fragment this manuscript inputs is generated from committed JSON records and
carries a blob digest of each source record and of the generator itself, which
a structural checker re-derives; a record regenerated without rerunning the
generator therefore fails rather than typesetting stale numbers.  The same
checker verifies balanced environments, references, citations, and table
column counts.  Krylov width is quoted only when its numerical cutoff passes a
support count certified to describe the same normalized pencil, rank, energy,
and condition number as the unpruned calculation.

The repository distinguishes \emph{exact}, \emph{oracle sampled},
\emph{finite sample}, and \emph{heuristic} evidence.  These labels prevent a
bound on one growth decision from becoming a claim about a final noisy
spectrum.  The operator-bank implementation builds on the representation and
finite-shot selection framework introduced previously
\cite{utama2026operatorcentric}.

\section{Conclusion}
\label{sec:conclusion}

DA-CASE fixes one reference and reconstructs projected overlap, Hamiltonian,
and observable matrices from one reusable Pauli bank.  On H$_4$, changing only
generator resolution reduces the retained bank from \(7371\) to \(2240\)
words while preserving the nine-dimensional subspace and energy to machine
precision.  A reference-aware certificate makes the narrower span
sector-safe for the declared reference.  A separate dyadic measurement layer
reduces \(913\) QWC settings to \(64\) fully commuting settings, with explicit
logical-CX cost.  In a finite-shot TFIM diagnostic, covariance-aware
allocation and mode-wise overlap regularization reduce tail failures while
sacrificing median accuracy.

The method is not the most accurate matched-\(M\) arm, its selection cache is
large, and its current evidence does not support scaling or hardware
superiority.  Its contribution is more modest: one bank can serve several
projected quantities, and the resulting trade can be audited without confusing
state contexts, settings, shots, depth, or retries.  That accounting is the
necessary starting point for an end-to-end comparison.

\begin{acknowledgments}
The authors thank the Department of Engineering Physics, Institut Teknologi
Bandung.
\end{acknowledgments}

\bibliographystyle{apsrev4-2}
\bibliography{references}

@article{mcclean2017qse,
  author = {McClean, Jarrod R. and Kimchi-Schwartz, Mollie E. and Carter, Jonathan and de Jong, Wibe A.},
  title = {Hybrid quantum-classical hierarchy for mitigation of decoherence and determination of excited states},
  journal = {Physical Review A},
  volume = {95},
  pages = {042308},
  year = {2017},
  doi = {10.1103/PhysRevA.95.042308}
}

@article{colless2018spectra,
  author = {Colless, J. I. and Ramasesh, V. V. and Dahlen, D. and Blok, M. S. and Kimchi-Schwartz, M. E. and McClean, J. R. and Carter, J. and de Jong, W. A. and Siddiqi, I.},
  title = {Computation of molecular spectra on a quantum processor with an error-resilient algorithm},
  journal = {Physical Review X},
  volume = {8},
  pages = {011021},
  year = {2018},
  doi = {10.1103/PhysRevX.8.011021}
}

@article{huggins2020nonorthogonal,
  author = {Huggins, William J. and Lee, Joonho and Baek, Unpil and O'Gorman, Bryan and Whaley, K. Birgitta},
  title = {A non-orthogonal variational quantum eigensolver},
  journal = {New Journal of Physics},
  volume = {22},
  pages = {073009},
  year = {2020},
  doi = {10.1088/1367-2630/ab867b}
}

@article{stair2020krylov,
  author = {Stair, Nicholas H. and Huang, Renke and Evangelista, Francesco A.},
  title = {A multireference quantum Krylov algorithm for strongly correlated electrons},
  journal = {Journal of Chemical Theory and Computation},
  volume = {16},
  pages = {2236--2245},
  year = {2020},
  doi = {10.1021/acs.jctc.9b01125}
}

@article{epperly2022theory,
  author = {Epperly, Ethan N. and Lin, Lin and Nakatsukasa, Yuji},
  title = {A theory of quantum subspace diagonalization},
  journal = {SIAM Journal on Matrix Analysis and Applications},
  volume = {43},
  pages = {1263--1290},
  year = {2022},
  doi = {10.1137/21M145954X}
}

@article{tkachenko2024davidson,
  author = {Tkachenko, Nikolay V. and Cincio, Lukasz and Boldyrev, Alexander I. and Tretiak, Sergei and Dub, Pavel A. and Zhang, Yu},
  title = {Quantum Davidson algorithm for excited states},
  journal = {Quantum Science and Technology},
  volume = {9},
  pages = {035012},
  year = {2024},
  doi = {10.1088/2058-9565/ad3a97}
}

@article{lee2024sampling,
  author = {Lee, Gwonhak and Lee, Dongkeun and Huh, Joonsuk},
  title = {Sampling error analysis in quantum Krylov subspace diagonalization},
  journal = {Quantum},
  volume = {8},
  pages = {1477},
  year = {2024},
  doi = {10.22331/q-2024-09-19-1477}
}

@article{zhang2024measurement,
  author = {Zhang, Zongkang and Wang, Anbang and Xu, Xiaosi and Li, Ying},
  title = {Measurement-efficient quantum Krylov subspace diagonalisation},
  journal = {Quantum},
  volume = {8},
  pages = {1438},
  year = {2024},
  doi = {10.22331/q-2024-08-13-1438}
}

@article{oleary2025partitioned,
  author = {O'Leary, Tom and Anderson, Lewis W. and Jaksch, Dieter and Kiffner, Martin},
  title = {Partitioned quantum subspace expansion},
  journal = {Quantum},
  volume = {9},
  pages = {1726},
  year = {2025},
  doi = {10.22331/q-2025-05-05-1726}
}

@article{nakagawa2024adaptqsci,
  author = {Nakagawa, Yuya O. and Kamoshita, Masahiko and Mizukami, Wataru and Sudo, Shotaro and Ohnishi, Yu-ya},
  title = {{ADAPT-QSCI}: Adaptive construction of an input state for quantum-selected configuration interaction},
  journal = {Journal of Chemical Theory and Computation},
  volume = {20},
  pages = {10817--10825},
  year = {2024},
  doi = {10.1021/acs.jctc.4c00846}
}

@article{umeano2025response,
  author = {Umeano, Chukwudubem and Jamet, Fran\c{c}ois and Lindoy, Lachlan P. and Rungger, Ivan and Kyriienko, Oleksandr},
  title = {Quantum subspace expansion approach for simulating dynamical response functions of Kitaev spin liquids},
  journal = {Physical Review Materials},
  volume = {9},
  pages = {034401},
  year = {2025},
  doi = {10.1103/PhysRevMaterials.9.034401}
}

@article{peruzzo2014vqe,
  author = {Peruzzo, Alberto and McClean, Jarrod and Shadbolt, Peter and Yung, Man-Hong and Zhou, Xiao-Qi and Love, Peter J. and Aspuru-Guzik, Al\'an and O'Brien, Jeremy L.},
  title = {A variational eigenvalue solver on a photonic quantum processor},
  journal = {Nature Communications},
  volume = {5},
  pages = {4213},
  year = {2014},
  doi = {10.1038/ncomms5213}
}

@article{grimsley2019adapt,
  author = {Grimsley, Harper R. and Economou, Sophia E. and Barnes, Edwin and Mayhall, Nicholas J.},
  title = {An adaptive variational algorithm for exact molecular simulations on a quantum computer},
  journal = {Nature Communications},
  volume = {10},
  pages = {3007},
  year = {2019},
  doi = {10.1038/s41467-019-10988-2}
}

@article{verteletskyi2020grouping,
  author = {Verteletskyi, Vladyslav and Yen, Tzu-Ching and Izmaylov, Artur F.},
  title = {Measurement optimization in the variational quantum eigensolver using a minimum clique cover},
  journal = {Journal of Chemical Physics},
  volume = {152},
  pages = {124114},
  year = {2020},
  doi = {10.1063/1.5141458}
}

@article{crawford2021efficient,
  author = {Crawford, Ophelia and van Straaten, Barnaby and Wang, Daochen and Parks, Thomas and Campbell, Earl and Brierley, Stephen},
  title = {Efficient quantum measurement of Pauli operators in the presence of finite sampling error},
  journal = {Quantum},
  volume = {5},
  pages = {385},
  year = {2021},
  doi = {10.22331/q-2021-01-20-385}
}

@article{miller2024hardware,
  author = {Miller, Daniel and Fischer, Laurin E. and Levi, Kyano and Kuehnke, Eric J. and Sokolov, Igor O. and Barkoutsos, Panagiotis Kl. and Eisert, Jens and Tavernelli, Ivano},
  title = {Hardware-Tailored Diagonalization Circuits},
  journal = {npj Quantum Information},
  volume = {10},
  pages = {122},
  year = {2024},
  doi = {10.1038/s41534-024-00901-1}
}

@article{efron1979bootstrap,
  author = {Efron, Bradley},
  title = {Bootstrap methods: Another look at the jackknife},
  journal = {The Annals of Statistics},
  volume = {7},
  pages = {1--26},
  year = {1979},
  doi = {10.1214/aos/1176344552}
}

@article{sun2018pyscf,
  author = {Sun, Qiming and Berkelbach, Timothy C. and Blunt, Nick S. and Booth, George H. and Guo, Sheng and Li, Zhendong and Liu, Junzi and McClain, James D. and Sayfutyarova, Elvira R. and Sharma, Sandeep and Wouters, Sebastian and Chan, Garnet Kin-Lic},
  title = {{PySCF}: the Python-based simulations of chemistry framework},
  journal = {WIREs Computational Molecular Science},
  volume = {8},
  pages = {e1340},
  year = {2018},
  doi = {10.1002/wcms.1340}
}

@article{knowles1989fcidump,
  author = {Knowles, Peter J. and Handy, Nicholas C.},
  title = {A determinant based full configuration interaction program},
  journal = {Computer Physics Communications},
  volume = {54},
  pages = {75--83},
  year = {1989},
  doi = {10.1016/0010-4655(89)90033-7}
}

@article{mcclean2020openfermion,
  author = {McClean, Jarrod R. and Rubin, Nicholas C. and Sung, Kevin J. and others},
  title = {{OpenFermion}: the electronic structure package for quantum computers},
  journal = {Quantum Science and Technology},
  volume = {5},
  pages = {034014},
  year = {2020},
  doi = {10.1088/2058-9565/ab8ebc}
}

@article{marzari2012wannier,
  author = {Marzari, Nicola and Mostofi, Arash A. and Yates, Jonathan R. and Souza, Ivo and Vanderbilt, David},
  title = {Maximally localized Wannier functions: Theory and applications},
  journal = {Reviews of Modern Physics},
  volume = {84},
  pages = {1419--1475},
  year = {2012},
  doi = {10.1103/RevModPhys.84.1419}
}

@article{pizzi2020wannier90,
  author = {Pizzi, Giovanni and Vitale, Valerio and Arita, Ryotaro and others},
  title = {{Wannier90} as a community code: new features and applications},
  journal = {Journal of Physics: Condensed Matter},
  volume = {32},
  pages = {165902},
  year = {2020},
  doi = {10.1088/1361-648X/ab51ff}
}

@article{aryasetiawan2004crpa,
  author = {Aryasetiawan, F. and Imada, M. and Georges, A. and Kotliar, G. and Biermann, S. and Lichtenstein, A. I.},
  title = {Frequency-dependent local interactions and low-energy effective models from electronic structure calculations},
  journal = {Physical Review B},
  volume = {70},
  pages = {195104},
  year = {2004},
  doi = {10.1103/PhysRevB.70.195104}
}

@article{georges1996dmft,
  author = {Georges, Antoine and Kotliar, Gabriel and Krauth, Werner and Rozenberg, Marcelo J.},
  title = {Dynamical mean-field theory of strongly correlated fermion systems and the limit of infinite dimensions},
  journal = {Reviews of Modern Physics},
  volume = {68},
  pages = {13--125},
  year = {1996},
  doi = {10.1103/RevModPhys.68.13}
}

@article{bharti2021iqae,
  author = {Bharti, Kishor and Haug, Tobias},
  title = {Iterative quantum-assisted eigensolver},
  journal = {Physical Review A},
  volume = {104},
  pages = {L050401},
  year = {2021},
  doi = {10.1103/PhysRevA.104.L050401}
}

@article{zheng2024adaptgcim,
  author = {Zheng, Muqing and Peng, Bo and Li, Ang and Yang, Xiu and Kowalski, Karol},
  title = {Unleashed from constrained optimization: quantum computing for quantum chemistry employing generator coordinate inspired method},
  journal = {npj Quantum Information},
  volume = {10},
  pages = {127},
  year = {2024},
  doi = {10.1038/s41534-024-00916-8}
}

@article{patel2026qsense,
  author = {Patel, Smik and Jayakumar, Praveen and Huang, Runjia and Zeng, Tao and Izmaylov, Artur F.},
  title = {Quantum seniority-based subspace expansion: Linear combinations of short-circuit unitary transformations for the electronic structure problem},
  journal = {Journal of Chemical Theory and Computation},
  volume = {22},
  number = {8},
  pages = {3937--3949},
  year = {2026},
  doi = {10.1021/acs.jctc.6c00017}
}

@misc{miura2026assqd,
  author = {Miura, Rinka},
  title = {Active sampling sample-based quantum diagonalization from finite-shot measurements},
  year = {2026},
  eprint = {2603.13536},
  archivePrefix = {arXiv},
  primaryClass = {quant-ph},
  doi = {10.48550/arXiv.2603.13536}
}

@article{utama2026operatorcentric,
  author = {Utama, Ginanjar and Dipojono, Hermawan Kresno},
  title = {Operator-centric Clifford algebra for variational eigensolvers and finite-shot adaptive selection},
  journal = {arXiv preprint arXiv:2607.17443},
  year = {2026},
  doi = {10.48550/arXiv.2607.17443}
}

@article{kanno2026qsci,
  author = {Kanno, Keita and Kohda, Masaya and Imai, Ryosuke and Koh, Sho and Mitarai, Kosuke and Mizukami, Wataru and Nakagawa, Yuya O.},
  title = {Quantum-Selected Configuration Interaction: classical diagonalization of Hamiltonians in subspaces selected by quantum computers},
  journal = {Physical Review Research},
  volume = {8},
  pages = {023268},
  year = {2026},
  doi = {10.1103/dmn4-snfx}
}

@misc{feniou2023overlapadapt,
  author = {Feniou, Cesar and Hassan, Muhammad and Traore, Diata and Giner, Emmanuel and Maday, Yvon and Piquemal, Jean-Philip},
  title = {Overlap-ADAPT-VQE: Practical Quantum Chemistry on Quantum Computers via Overlap-Guided Compact Ans\"atze},
  year = {2023},
  eprint = {2301.10196},
  archivePrefix = {arXiv},
  primaryClass = {quant-ph},
  doi = {10.48550/arXiv.2301.10196}
}

@misc{gaberle2026critical,
  author = {Gaberle, Cedric and Jattana, Manpreet S.},
  title = {A Critical Assessment of the Sample-Based Quantum Diagonalization for Heisenberg and Hubbard Models},
  year = {2026},
  eprint = {2605.02494},
  archivePrefix = {arXiv},
  primaryClass = {quant-ph},
  doi = {10.48550/arXiv.2605.02494}
}

\end{document}